\documentclass{article}
\usepackage{amsfonts}
\usepackage{makeidx}
\usepackage{amsmath}
\usepackage{amssymb}
\usepackage{graphicx}

\newtheorem{theorem}{Theorem}

\input{tcilatex}

\begin{document}

\title{{\Large \textbf{Lie and Noether symmetries of geodesic equations and
collineations}}}
\author{Michael Tsamparlis\thanks{%
Email: mtsampa@phys.uoa.gr} \  and Andronikos Paliathanasis\thanks{%
Email: anpaliat@phys.uoa.gr} \\
{\small \textit{Faculty of Physics, Department of
Astronomy-Astrophysics-Mechanics,}}\\
{\small \textit{\ University of Athens, Panepistemiopolis, Athens 157 83,
GREECE}}}
\date{}
\maketitle

\begin{abstract}
The Lie symmetries of the geodesic equations in a Riemannian space are
computed in terms of the special projective group and its degenerates
(affine vectors, homothetic vector and Killing vectors) of the metric. The
Noether symmetries of the same equations are given in terms of the
homothetic and the Killing vectors of the metric. It is shown that the
geodesic equations in a Riemannian space admit three linear first integrals
and two quadratic first integrals. We apply the results in the case of
Einstein spaces, the Schwarzschild spacetime and the Friedman Robertson
Walker spacetime. In each case the Lie and the Noether symmetries are
computed explicitly together with the corresponding linear and quadratic
first integrals.
\end{abstract}

Keywords:\newline
Geodesics, General Relativity, Classical Mechanics, Collineations, Lie
Symmetries, Projective Collineations, Noether Symmetries, First Integrals.%
\newline

PACS - numbers: 2.40.Hw, 4.20.-q, 4.20.Jb, 04.20.Me, 03.20.+i, 02.40.Ky

\section{Introduction}

\label{Introduction}

Geometrically the Lie symmetries of a system of differential equations are
understood as automorphisms which preserve the solution curves. In a space
with a linear connection there is an inherent system of differential
equations defined by the paths (or autoparallels) of the connection. It is
well known that these curves (as a set) are preserved under the projective
automorphisms of the space. Therefore it is reasonable one to expect a
relation between the projective collineations of the space and the Lie
symmetries of the system of differential equations of the paths.

A special case of the above scenario occurs in a Riemannian space in which
case the paths are the (affinely or not) parameterized geodesics determined
by the metric. In this case one expects a close relation to exist between
the geodesic equations and the projective collineations (PC) of the metric
or their degenerates affine collineations (AC), homothetic Killing vector
(HKV) and Killing vectors (KV).

Indeed this topic has been discussed extensively in the literature. Classic
is the work of Katzin and Levin (\cite{Katzin Levine 1972 Poland}, \cite%
{Katzin Levine JMP 15 1974}, \cite{Katzin Levine JMP 17 1976},\cite{Katzin
Levine JMP 22 1981}). Important contributions have also been done by Aminova
(\cite{Aminova 1978}, \cite{Aminova 1994}, \cite{Aminova 1995}, \cite%
{Aminova 2000}), Prince and Crampin \cite{Prince Crampin (1984) 1} and many
others. More recent is the work of Feroze et al \cite{Feroze Mahomed Qadir}
where the case of maximally symmetric spaces of low dimension it is
discussed .

Furthermore, because the geodesic equations follow from the variation of the
geodesic Lagrangian defined by the metric and due to the fact that the
Noether symmetries are a subgroup of the Lie group of Lie symmetries of
these equations, one should expect a relation / identification of the
Noether symmetries of this Lagrangian with the projective collineations of
the metric or with its degenerates. Recent work in this direction has been
by Bokhari et all (\cite{Bokhari 2006 Int Jour Theor Phys},\cite{Bokhari
2007 Gen Rel Grav}) in which the relation of the Noether symmetries with the
KVs of some special spacetimes it is discussed.

In the present paper we give a complete answer to both topics mentioned
above. In Section \ref{section 2} we give a brief introduction concerning
the autoparallels of a symmetric connection. In Section \ref{section 3} we
determine the conditions for the Lie symmetries in covariant form and relate
them with the special projective symmetries of the connection. A similar
result has been obtained by Prince and Crampin in \cite{Prince Crampin
(1984) 1} using the bundle formulation of second order ordinary differential
equations (ODE). In Section \ref{section 4} we apply these conditions in the
special case of Riemannian spaces and in Theorem \ref{Theorem 1} we give the
Lie symmetry vectors in terms of the special projective collineations of the
metric and their degenerates. In Section \ref{Noether symmetries} we give
the second result of this work, that is Theorem \ref{Theorem Noether
symmetries of GEqs}, which relates the Noether symmetries of the geodesic
Lagrangian defined by the metric with the homothetic algebra of the metric
and comment on the results obtained so far in the literature. Finally in
Section \ref{section 5} we apply the results to various cases and eventually
we give the Noether symmetries and the associated conserved quantities of
the Friedman Robertson Walker (FRW) spacetimes.

\section{Preliminary results}

\label{section 2}

Consider a $C^{\infty }$ manifold $M$ of dimension $n$, endowed with a $%
\Gamma _{jk}^{i}$ symmetric\footnote{%
The coefficients $\Gamma _{jk}^{i}$ in general are not symmetric in the
lower indices. In the autoparallel equation (\ref{PCA.1}) the antisymmetric
part of $\Gamma _{\lbrack jk]}^{i}$ (the torsion) does not play a role.}
connection. In a local coordinate system $\{x^{i}|i=1,\ldots ,n\}$ the
connection $\Gamma _{jk}^{i}\partial _{i}=\nabla _{j}\partial _{k}$ and the
autoparallels of the connection are defined by the requirement: 
\begin{equation}
\ddot{x}^{i}(t)+\Gamma _{jk}^{i}(x(t))\dot{x}^{j}(t)\dot{x}^{k}(t)=\phi (t)%
\dot{x}^{i}(t)\;,\quad i=1,\ldots ,n,  \label{PCA.1}
\end{equation}%
where $t$ is a parameter along the paths. When $\phi $ vanishes, we say that
the autoparallel is affinely parameterized and in this case $t$ is called an
affine parameter, that is one has:%
\begin{equation}
\ddot{x}^{i}(t)+\Gamma _{jk}^{i}(x(t))\dot{x}^{j}(t)\dot{x}^{k}(t)=0,\quad
i=1,\ldots ,n.  \label{PCA.6}
\end{equation}

If $\mathbf{X}=X^{a}\partial _{a}$ is a vector field on the manifold the
following identity holds (see Yano \cite{Yano 1956} eqn. (2.16)):%
\begin{equation}
\mathcal{L}_{X}\Gamma _{jk}^{i}=X_{,jk}^{i}+\Gamma
_{jk,l}^{i}X^{l}+X_{,k}^{l}\Gamma _{lj}^{i}+X_{,j}^{l}\Gamma
_{lk}^{i}-X_{,l}^{i}\Gamma _{jk}^{l}  \label{PCA.2}
\end{equation}

We say that the vector field $\mathbf{X}$ is an Affine Collineation (AC)
iff: 
\begin{equation}
\mathcal{L}_{X}\Gamma _{jk}^{i}=0.  \label{PCA.3}
\end{equation}%
In flat space equation (\ref{PCA.3}) implies the condition:%
\begin{equation}
X_{a,bc}=0  \label{PCA.4}
\end{equation}%
the solution of which is: 
\begin{equation}
X_{a}=B_{ab}x^{b}+C_{a}  \label{PCA.5}
\end{equation}%
where $B_{ab}$ and $C_{a}$ are constants. The geometric property/definition
of affine collineations is that they preserve the set of autoparallels (i.e.
paths) of the connection together with their affine parametrization (that
is, by an affine symmetry an affinely parameterized autoparallel goes over
to an affinely parameterized autoparallel of the same connection). From (\ref%
{PCA.5}) we infer that in an n-dimensional space there are at most $%
n+n^{2}=n(n+1)$ ACs and, when this is the case, it can be shown that the
space is flat.

We say that a vector field $\mathbf{X}\;$is a projective collineation of the
connection if there exists a one form $\omega _{i}$ such that the following
condition holds (\cite{Yano 1956}, \cite{Scouten 1954}): 
\begin{equation}
\mathcal{L}_{X}\Gamma _{jk}^{i}=\omega _{j}\delta _{k}^{i}+\omega _{k}\delta
_{j}^{i}.  \label{PCA.7}
\end{equation}

In a Riemannian space the form $\omega ^{i}$ is closed, that is, there
exists a function $f(x^{i}),$ called the projective function, such that:%
\begin{equation}
\mathcal{L}_{X}\Gamma _{jk}^{i}=f_{,j}\delta _{k}^{i}+f_{,k}\delta _{j}^{i}.
\label{PCA.7b}
\end{equation}

In flat space condition (\ref{PCA.7b}) implies that:%
\begin{equation}
X_{a,b}=B_{ab}+(A_{c}x^{c})g_{ab}+C_{b}x_{a}.  \label{PCA.7a}
\end{equation}%
which has the solution: 
\begin{equation}
X_{a}=B_{ab}x^{b}+(A_{b}x^{b})x_{a}+C_{a}  \label{PCA.8}
\end{equation}%
where again the various coefficients are constants. In an n-dimensional
space there are at most $n^{2}+n+n=n(n+2)$ projective collineations of the
connection and when this is the case, it can be shown that the space is
flat. This holds in any space irrespective of the signature of the metric
and the (finite) dimension of space. In case that the function $f$ satisfies
the condition $f_{;ij}=0,$ that is, $f_{,i}$ is a gradient KV, the
projective collineation is called special.

The geometric property/definition of the (proper) projective collineations
is that they preserve the set of autoparallels, but they do not preserve the
parametrization.

\section{Lie point symmetries of the autoparallel equations}

\label{section 3} We write the system of ODEs (\ref{PCA.1}) in the form $%
\ddot{x}^{i}=\omega ^{i}(x,\dot{x},t)$ where: 
\begin{equation}
\omega ^{i}(x,\dot{x},t)=-\Gamma _{jk}^{i}(x)\dot{x}^{j}\dot{x}^{k}-\phi (x)%
\dot{x}^{i}.  \label{PCA.14}
\end{equation}%
The associated linear operator defined by this system of ODEs is: 
\begin{equation}
\mathbf{A}=\frac{\partial }{\partial t}+\dot{x}^{i}\frac{\partial }{\partial
x^{i}}+\omega ^{i}(t,x^{j},\dot{x}^{j})\frac{\partial }{\partial \dot{x}^{i}}%
.  \label{PCA.16}
\end{equation}%
The condition for a Lie point symmetry for the system of equations is \cite%
{Stephani book ODES}: 
\begin{equation}
\lbrack X^{[1]},\mathbf{A}]=\lambda (x^{a})\mathbf{A}  \label{PCA.17}
\end{equation}%
where $X^{[1]}$ is the first prolongation of the symmetry vector $X=\xi
(t,x)\partial _{t}+\eta ^{i}(t,x)\partial _{x^{i}}$ defined as follows:

\begin{equation}
X^{[1]}=\xi (t,x,\dot{x})\partial _{t}+\eta ^{i}(t,x,\dot{x})\partial
_{x^{i}}+G^{[1]i}\partial _{\dot{x}^{i}}  \label{PCA.12}
\end{equation}%
and $G^{[1]i}$ are the coefficients of the first prolongation given by the
formula: 
\begin{equation}
G^{[1]i}=\frac{d}{dt}\eta ^{i}-\dot{x}^{i}\frac{d}{dt}\xi =\eta
_{,t}^{i}+\eta _{,j}^{i}\dot{x}^{j}-\xi _{,t}\dot{x}^{i}-\xi _{,j}\dot{x}^{i}%
\dot{x}^{j}\;.  \notag
\end{equation}%
It is a standard result \cite{Stephani book ODES} that (\ref{PCA.17}) leads
to the following three conditions: 
\begin{align}
-\mathbf{A}\xi & =\lambda  \label{PCA.18} \\
G^{[1]i}& =\mathbf{A}\eta ^{i}-\dot{x}^{i}\mathbf{A}\xi  \label{PCA.19} \\
X^{[1]}(\omega ^{i})-\mathbf{A}(G^{[1]{i}})& =-\omega ^{i}\mathbf{A}\xi .
\label{PCA.20}
\end{align}

For any function, $f(t,x^{i})$ , $\mathbf{A}f=df/dt=f_{,t}+f_{,i}\dot{x}^{i}$
is the total derivative of $f.$ Using this result we write the symmetry
conditions as follows:%
\begin{align}
\lambda & =-\frac{d\xi }{dt}  \label{PCA.21} \\
G^{[1]i}& =\frac{d\eta ^{i}}{dt}-\dot{x}^{i}\frac{d\xi }{dt}  \label{PCA.22}
\\
X^{[1]}(\omega ^{i})-A(G^{[1]{i}})& =-\omega ^{i}\frac{d\xi }{dt}.
\label{PCA.23}
\end{align}

We note that the second condition (\ref{PCA.22}) defines the first
prolongation $G^{[1]i}.$ The first equation (\ref{PCA.21}) gives the factor $%
\lambda .$ Therefore the essential symmetry condition is equation (\ref%
{PCA.23}).

To compute the symmetry condition we have to compute the quantities $%
X^{[1]}(\omega ^{i})$ and $A(G^{[1]{i}})-\omega ^{i}\frac{d\xi }{dt}$ taking
into consideration (\ref{PCA.22}) and (\ref{PCA.14}). The result of this
formal calculation is: 
\begin{align}
X^{[1]}(\omega ^{i})& =(\xi \partial _{t}+\eta ^{i}\partial
_{x^{i}}+G^{[1]i}\partial _{\dot{x}^{i}})(-\Gamma _{jk}^{i}(x)\dot{x}^{j}%
\dot{x}^{k}-\phi (x)\dot{x}^{i})  \notag \\
& =-\eta _{,t}^{i}\phi +(-\xi \phi _{,t}\delta _{j}^{i}-\phi _{,k}\eta
^{k}\delta _{j}^{i}-\eta _{,t}^{k}\Gamma _{kj}^{i}-\eta _{,t}^{k}\Gamma
_{jk}^{i}-\phi \eta _{,j}^{i}+\phi \xi _{,t}\delta _{j}^{i})\dot{x}^{j} 
\notag \\
& \qquad +(-\xi \Gamma _{(kj),t}^{i}-\eta ^{l}\Gamma _{(kj),l}^{i}-\eta
_{,k}^{l}\Gamma _{(lj)}^{i}-\eta _{,k}^{l}\Gamma _{(jl)}^{i}+\phi \xi
_{,k}\delta _{j}^{i}+2\xi _{,t}\Gamma _{(kj)}^{i})\dot{x}^{j}\dot{x}^{k}
\label{PCA.30} \\
& \qquad +\xi _{,(k}\Gamma _{jl)}^{i}\dot{x}^{j}\dot{x}^{k}\dot{x}^{l}. 
\notag
\end{align}%
\begin{align*}
A(G^{[1]i})-\omega ^{i}\frac{d\xi }{dt}& =\eta _{,tt}^{i}+(2\eta
_{,tj}^{i}+\phi \eta _{,j}^{i}-2\phi \xi _{,t}\delta _{j}^{i}-\xi
_{,tt}\delta _{j}^{i})\dot{x}^{j} \\
& \qquad +(\eta _{,(jk)}^{i}-2\xi _{,t(j}\delta _{k)}^{i}-\eta
_{,l}^{i}\Gamma _{(jk)}^{l}+2\xi _{,t}\Gamma _{(jk)}^{i}-2\phi \xi
_{(,j}\delta _{k)}^{i})\dot{x}^{j}\dot{x}^{k} \\
& \qquad +(\xi _{,(j}\Gamma _{kl)}^{i}+\xi _{,m}\Gamma _{(kl}^{m}\delta
_{j)}^{i}-\xi _{,(jk}\delta _{l)}^{i})\dot{x}^{j}\dot{x}^{k}\dot{x}^{l}.
\end{align*}%
Substituting into the symmetry condition (\ref{PCA.23}) and collecting terms
of the same order in $\dot{x}^{j}$ we find the following equations: ($%
i=1,\ldots ,n$):\newline
$(\dot{x})^{0}$ terms: 
\begin{equation}
\eta _{,tt}^{i}+\eta _{,t}^{i}\phi =0  \label{PCA.31}
\end{equation}%
$(\dot{x})^{1}$ terms: 
\begin{equation}
\xi _{,tt}\delta _{j}^{i}-\xi \phi _{,t}\delta _{j}^{i}-2[\eta
_{,tj}^{i}+\eta _{,t}^{k}\Gamma _{(kj)}^{i}]-[\phi \xi _{,t}+\phi _{,k}\eta
^{k}]\delta _{j}^{i}=0  \label{PCA.32}
\end{equation}%
$(\dot{x})^{2}$ terms$:$%
\begin{equation*}
(-\eta _{,(jk)}^{i}-\eta ^{l}\Gamma _{(jk),l}^{i}-\eta _{,k}^{l}\Gamma
_{lj}^{i}-\eta _{,k}^{l}\Gamma _{jl}^{i}+\eta _{,l}^{i}\Gamma
_{jk}^{l})+2\xi _{,t(j}\delta _{k)}^{i}-2\phi \xi _{(,j}\delta _{k)}^{i}-\xi
\Gamma _{(kj),t}^{i}=0\mbox{\qquad
}\Rightarrow
\end{equation*}

\begin{equation}
\mathcal{L}_{\mathbf{\eta}}\Gamma_{(jk)}^{i}=-2\phi\xi_{(,j}\delta_{k)}^{i}+%
\xi\Gamma_{(kj),t}^{i}+2\xi_{,t(j}\delta_{k)}^{i}  \label{PCA.34}
\end{equation}
$(\dot{x})^{3}$ terms: 
\begin{equation}
(\xi_{,(jk}-\xi_{,|e|}\Gamma_{(jk}^{e})\delta_{l)}^{i}=0\;.  \label{PCA.35}
\end{equation}

Define the quantity: 
\begin{equation}
\Phi =\xi _{,t}-\phi \xi .  \label{PCA.36}
\end{equation}%
Then condition (\ref{PCA.34}) is written (note that $\phi ,_{i}=0)$:%
\begin{equation}
\mathcal{L}_{\mathbf{\eta }}\Gamma _{(jk)}^{i}=2\Phi _{(,j}\delta
_{k)}^{i}-\xi \Gamma _{(kj),t}^{i}.  \label{PCA.37}
\end{equation}%
If we consider the vector\textbf{\ }$\mathbb{\xi }\mathbf{=}\xi \partial
_{t} $ (which does not have components along $\partial _{i}$), we find that:%
\begin{equation*}
\mathcal{L}_{\mathbf{\xi }}\Gamma _{(jk)}^{i}=\xi \Gamma _{(kj),t}^{i}.
\end{equation*}%
Hence (\ref{PCA.37}) is written as: 
\begin{equation}
\mathcal{L}_{\mathbf{X}}\Gamma _{(jk)}^{i}=2\Phi _{(,j}\delta _{k)}^{i},
\label{PCA.38}
\end{equation}%
where $\mathbf{X}=\mathbb{\xi }\mathbf{+}\mathbb{\eta }\mathbf{=}\xi
\partial _{t}+\eta ^{i}(t,x)\partial _{x^{i}}.$ We note that this condition
is precisely condition (\ref{PCA.7}) for a projective collineation of the
connection $\Gamma _{(jk)}^{i}$ along the symmetry vector $\mathbf{X}$ and
with projecting function $\Phi $. Concerning the other conditions we note
that (\ref{PCA.32}) can be written in covariant form (relevant to the
indices $a=1,2,\ldots,n)$ as follows: 
\begin{equation}
\Phi _{,t}\delta _{j}^{i}-2\eta _{,t\mid j}^{i}=0  \label{PCA.39}
\end{equation}%
where $\eta _{,t\mid j}^{i}=\eta _{,tj}^{i}+\eta _{,t}^{k}\Gamma _{(kj)}^{i}$
is the covariant derivative of the vector $\eta _{,t}^{i}$ with respect to $%
\Gamma _{(kj)}^{i}$. Similarly condition (\ref{PCA.35}) can be written as%
\begin{equation*}
\xi _{\mid (jk}\delta _{l)}^{i}=0.
\end{equation*}%
Contracting on the indices $i$ and $j$ we find the final form:%
\begin{equation}
\xi _{\mid (jk)}=0.  \label{PCA.40}
\end{equation}%
This implies that $\xi ,_{{i}}$ is a gradient KV of the metric of the space $%
\{x^{i}\}$.

Condition (\ref{PCA.31}) is obviously in covariant form wrt the indices $a.$

The Lie symmetries of the autoparallel equations (\ref{PCA.14}) (not
necessarily affinely parameterized) for a general connection defined on a $%
C^{\infty }$ manifold are given by the following covariant\footnote{%
These are covariant equations because, if we consider the connection in the
augmented $n+1$ space $\{x^{i},t\}$, all components of $\Gamma $ which
contain an index along the direction of $t$ vanish. Therefore the partial
derivatives wrt $t$ can be replaced with a covariant derivative wrt $t.$}
equations:%
\begin{align}
\eta _{,tt}^{i}+\eta _{,t}^{i}\phi & =0  \label{PCA.51} \\
\xi _{\mid (jk)}& =0  \label{PCA.52} \\
\Phi _{,t}\delta _{j}^{i}-2\eta _{,t\mid j}^{i}& =0  \label{PCA.53} \\
\mathcal{L}_{\mathbf{X}}\Gamma _{(jk)}^{i}& =2\Phi _{(,j}\delta _{k)}^{i},
\label{PCA.54}
\end{align}%
where the Lie symmetry vector $\mathbf{X}=\mathbb{\xi }\mathbf{+}\mathbb{%
\eta }\mathbf{=}\xi (t,x)\partial _{t}+\eta ^{i}(t,x)\partial _{x^{i}}.$ In
particular in the $n-$dimensional space $\{x^{i}\}$ the vector $\xi _{,i}$
is a gradient Killing vector and the vectors $\eta ^{i}\emph{(t,x)}\partial
_{x^{i}}$ are projective collineations, whereas in the jet space $%
\{t,x^{i}\} $ the Lie symmetry vector $X$\ is an affine collineation.

In the following we restrict our considerations to the case of Riemannian
connections, that is, the $\Gamma_{jk}^{i}$ are symmetric and the covariant
derivative of the metric vanishes.

\section{Calculation of the Lie symmetry vectors for a Riemannian connection}

\label{section 4}

We compute the Lie symmetry vectors for the case of affine parametrization ($%
\phi=0\ )$ and the assumption $\Gamma_{jk,t}^{i}=0$ \ i.e. the $\Gamma
_{jk}^{i}$ are independent of the parameter $t.$ The later is a logical
assumption because the $\Gamma_{jk}^{i}$ are computed in terms of the metric
which does not depend on the affine parameter $t$. Under these assumptions
the symmetry conditions (\ref{PCA.51}) - (\ref{PCA.54}) read:%
\begin{align}
\eta_{,tt}^{i} & =0  \label{PCA.55} \\
\xi_{\mid(jk)} & =0  \label{PCA.56} \\
\xi_{,tt}\delta_{j}^{i}-2\eta_{,t\mid j}^{i} & =0  \label{PCA.57} \\
\mathcal{L}_{\mathbf{\eta}}\Gamma_{jk}^{i} & =2\xi,_{t(j}\delta_{k)}^{i}.
\label{PCA.58}
\end{align}

The solution of this system of equations is given in the following Theorem.
The actual computations are given in the Appendix.

\begin{theorem}
\label{Theorem 1} The Lie symmetry vector $\mathbf{X}=\xi (t,x)\partial
_{t}+\eta ^{i}(t,x)\partial _{x^{i}}$ of the equations of geodesics (\ref%
{PCA.6}) in a Riemannian space involves all symmetry vectors, that is, KVs,
HKVs, ACs and special PCs as follows:\newline
A. The metric admits gradient KVs. Then

a. The function%
\begin{equation}
\xi (t,x)=\frac{1}{2}\left( G_{J}S^{J}+M\right) t^{2}+\left[ E_{J}S^{J}+K%
\right] t+F_{J}S^{J}+L,  \label{PCA.70b}
\end{equation}%
where $G_{J},M,b,K,F_{J}$ and $L$ \ are constants and the index $J\ $runs
along the number of gradient KVs

b. The vector%
\begin{equation}
\eta ^{i}(t,x)=A^{i}(x)t+B^{i}(x)+D^{i}(x)  \label{PCA.70d}
\end{equation}%
where the vector $A^{i}(x)$ is a gradient HKV with conformal factor $\psi =%
\frac{1}{2}\left( G_{J}S^{J}+M\right) $ (if it exists), $D^{i}(x)$ is a
non-gradient KV of the metric and $B^{i}(x)$ is either a special projective
collineation with projection function $E_{J}S^{J}(x)$ or an AC and $E_{J}=0$
in (\ref{PCA.70b}).

\noindent B. The metric does not admit gradient KVs. Then

a. The function%
\begin{equation}
\xi (t,x)=\frac{1}{2}Mt^{2}+Kt+L  \label{PCA.70cc}
\end{equation}

b. The vector%
\begin{equation}
\eta ^{i}(t,x)=A^{i}(x)t+B^{i}(x)+D^{i}(x),  \label{PCA.70e}
\end{equation}%
where $A^{i}(x)$ is a gradient HKV with conformal factor $\psi =\frac{1}{2}%
M, $ $D^{i}(x)$ is a non-gradient KV\ of the metric and $B^{i}(x)$ is an AC.
If in addition the metric does not admit gradient HKV, then%
\begin{align}
\xi (t)& =Kt+L  \label{PCA.70g} \\
\eta ^{i}(x)& =B^{i}(x)+D^{i}(x).  \label{PCA.70f}
\end{align}
\end{theorem}

\section{The Noether symmetries of the geodesic Lagrangian}

\label{Noether symmetries}

In a Riemannian space the equations of geodesics (\ref{PCA.3}) are produced
from the geodesic Lagrangian:%
\begin{equation}
L=\frac{1}{2}g_{ij}\dot{x}^{i}\dot{x}^{j}~.  \label{NS.6}
\end{equation}%
A\ vector field $\mathbf{X}=\xi \left( t,x^{k}\right) \partial _{t}+\eta
^{i}\left( t,x^{k}\right) \partial _{x^{i}}$ is a Noether symmetry of this
Lagrangian if there exists a smooth function $f(t,x^{i})$ such that \cite%
{Stephani book ODES}%
\begin{equation}
X^{\left[ 1\right] }L+\frac{d\xi }{dt}L=\frac{df}{dt},  \label{NS.7}
\end{equation}%
where $X^{\left[ 1\right] }=\xi \left( t,x^{k}\right) \partial _{t}+\eta
^{i}\left( t,x^{k}\right) \partial _{x^{i}}+\left( \frac{d\eta ^{i}}{dt}-%
\dot{x}^{i}\frac{d\xi }{dt}\right) \partial _{\dot{x}^{i}}$ is the first
prolongation of $\mathbf{X}.$ We compute:%
\begin{equation*}
X^{\left[ 1\right] }L=\frac{1}{2}\left( \eta ^{k}g_{ij,k}\dot{x}^{i}\dot{x}%
^{j}+2\frac{d\eta ^{k}}{dt}g_{ik}\dot{x}^{i}-2\dot{x}^{i}\dot{x}^{j}\frac{%
d\xi }{dt}g_{ij}\right) .
\end{equation*}%
Replacing the total derivatives in the rhs 
\begin{align*}
\frac{d\xi }{dt}& =\xi _{,t}+\dot{x}^{k}\xi _{,k} \\
\frac{d\eta ^{i}}{dt}& =\eta _{,t}^{i}+\dot{x}^{k}\eta _{,k}^{i}
\end{align*}%
we find that%
\begin{equation*}
X^{\left[ 1\right] }L=\frac{1}{2}\left( \eta ^{k}g_{ij,k}\dot{x}^{i}\dot{x}%
^{j}+2\eta _{,t}^{i}g_{ij}\dot{x}^{j}+\eta _{,r}^{i}g_{ik}\dot{x}^{k}\dot{x}%
^{r}+\eta _{,r}^{i}g_{kj}\dot{x}^{k}\dot{x}^{r}-2\xi _{,t}g_{ij}\dot{x}^{i}%
\dot{x}^{j}-2\xi _{,k}g_{ij}\dot{x}^{i}\dot{x}^{j}\dot{x}^{k}\right) .
\end{equation*}%
The term%
\begin{equation*}
\frac{d\xi }{dt}L=\frac{1}{2}\left( \xi _{,t}+\dot{x}^{k}\xi _{,k}\right)
g_{ij}\dot{x}^{i}\dot{x}^{j}.
\end{equation*}%
Finally the Noether symmetry condition (\ref{NS.7}) is

\begin{equation*}
-2f_{,t}+\biggl[2\eta _{,t}^{i}g_{ij}-2f_{,i}\biggr ]\dot{x}^{j}-\xi
_{,k}g_{ij}\dot{x}^{i}\dot{x}^{j}\dot{x}^{k}+\left[ \eta ^{k}g_{ij,k}+\eta
_{,i}^{k}g_{ik}+\eta _{,i}^{k}g_{kj}-g_{ij}\xi _{,t}\right] \dot{x}^{i}\dot{x%
}^{j}=0.
\end{equation*}

This relation is an identity hence the coefficient of each power of $\dot{x}%
^{j}$ must vanish. This results in the equations: 
\begin{align}
\dot{x}^{i}\dot{x}^{j}\dot{x}^{k}& :\xi _{,k}=0  \label{NS.8} \\
\dot{x}^{i}\dot{x}^{j}& :L_{\eta }g_{ij}=2\left( \frac{1}{2}\xi _{,t}\right)
g_{ij}  \label{NS.9} \\
\dot{x}^{i}& :\eta _{,t}^{,i}g_{ij}=f_{,i}  \label{NS.10} \\
\left( \dot{x}^{i}\right) ^{0}& :f_{,t}=0  \label{NS.11}
\end{align}

\bigskip Condition (\ref{NS.8}) gives $\xi_{,k}=0 \Rightarrow \xi=\xi\left(
t\right) $.

Condition (\ref{NS.11}) implies $f\left( x^{k}\right) $ and then condition (%
\ref{NS.8}) gives that $\eta ^{i}$ is of the form:%
\begin{equation}
\eta _{i}=f_{,i}t+K_{i}(x^{j}).  \label{NS.12}
\end{equation}%
Then from (\ref{NS.9}) follows that $\xi _{,t}$ must be at most linear in $%
t. $ Hence $\xi (t)$ must be at most a function of $t^{2}.$ Furthermore from
(\ref{NS.9}) follows that $\eta ^{i}$ is at most a CKV with conformal factor 
$\psi _{H}=\frac{1}{2}(At+B),$ where $A,B$ are constants. We consider
various cases.

Case 1: Suppose $\xi =$constant=$C_{1}$. Then $\eta ^{i}$ is a KV\ of the
metric which is independent of $t.$ This implies that either $f_{,i}=0$ and $%
f=$constant $=A=0$ or that $f_{,i}$ is a gradient KV. In this case the
Noether symmetry vector is:%
\begin{equation*}
X^{i}=C_{1}\partial _{t}+g^{ij}\left( f_{,j}t+K_{j}(x^{r})\right) ,
\end{equation*}%
where $K^{i}$ is a non-gradient KV of $g_{ij}.$

Case 2: Suppose $\xi=2t.$ Then $\eta^{i}$ is a HKV of the metric $g_{ij}$
with homothetic factor $1$. Then $\eta_{i}=H_{i}(x^{j})$ , $%
f_{,i}=0\Rightarrow f=$constant$=0$ where $H^{i}$ is a HKV of $g_{ij}$ with
homothetic factor $\psi,$ not necessarily a gradient HKV. In this case the
Noether symmetry vector is:%
\begin{equation*}
X^{i}=2\psi t\partial_{t}+H^{i}(x^{r}).
\end{equation*}

Case 3: $\xi (t)=t^{2}.$ Then $\eta ^{i}$ is a HKV of the metric $g_{ij}$
(the variable $t$ cancels) with homothetic factor $1.$ Again $f_{,i}$ is a
gradient HKV with homothetic factor $\psi $ and the Noether symmetry vector
is%
\begin{equation*}
X^{i}=\psi t^{2}\partial _{t}+g^{ij}f_{,j}t.
\end{equation*}

Therefore we have the result.

\begin{theorem}
\label{Theorem Noether symmetries of GEqs}

The Noether Symmetries of the geodesic Lagrangian follow from the KVs and
the HKV of the metric $g_{ij}$ as follows:%
\begin{align}
\xi (t)& =C_{3}\psi t^{2}+2C_{2}\psi t+C_{1}  \label{NS.13} \\
\eta ^{i}&
=C_{J}S^{J,i}+C_{I}KV^{Ii}+C_{IJ}tS^{J,i}+C_{2}H^{i}(x^{r})+C_{3}t(GHV)^{i}
\label{NS.14} \\
f(x^{i})& =C_{1}+C_{2}+C_{I}+C_{J}+\left[ C_{IJ}S^{J}\right] +C_{3}\left[ GHV%
\right],  \label{NS.15}
\end{align}%
where $S^{J,i}$ are the $C_{J}$ gradient KVs, $KV^{Ii}$ are the $C_{I}$
non-gradient KVs, $H^{i}$ is a HKV\ not necessarily gradient and $(GHV)^{i}$
is the gradient HKV (if it exists) of the metric $g_{ij}$.
\end{theorem}

The importance of Theorem \ref{Theorem Noether symmetries of GEqs} is that
one is able to compute the Lie symmetries and the Noether symmetries of the
geodesic equations in a Riemannian space by computing the corresponding
collineation vectors avoiding the cumbersome formulation of the Lie symmetry
method. It is also possible to use the inverse approach and prove that a
space does not admit KVs, HKVs, ACs and special PCs by using the
calculational approach of the Lie symmetry method (assisted with algebraic
manipulation programmes)\ and avoid the hard approach of Differential
Geometric methods. In section \ref{section 5} we demonstrate the use of the
above results.

\section{First Integrals of the geodesic equations and collineations}

Consider a Riemannian space with metric $g_{ij}$. As we have shown in
Theorem \ref{Theorem 1}, the Lie symmetries of the geodesic equations of the
metric coincide with the KVs, the HKV, the ACs and special PCs of the metric 
$g_{ij}$ (if they are admitted). For each Noether symmetry one has the first
integral 
\begin{equation}
\mathbf{X}^{[1]}L+\frac{d\xi }{dt}L=\frac{df}{dt},  \label{FOIG.0}
\end{equation}%
where, as before, $X^{[1]}$ is the first prolongation of $\mathbf{X}.$ In
this section we study the relationship between first integrals and
corresponding conserved quantities of Noether symmetries.

We recall first some well-known definitions and results\cite{Katzin Levine
1972 Poland}.

Consider the geodesic with tangent vector $\lambda ^{k}=$ $\frac{dx^{i}}{ds}$
where $s$ is an affine parameter along the geodesic. An $m$th order First
Integral of the geodesic is a tensor quantity $A_{r_{1}...r_{m}}$ such that:%
\begin{equation}
A_{r_{1}...r_{m}}\lambda ^{r_{1}}...\lambda ^{r_{m}}=\text{constant}
\label{FOIG.1}
\end{equation}%
or equivalently (because $\lambda _{i;j}=0)$:%
\begin{equation}
P\left\{ A_{r_{1}...r_{m};k}\right\} \lambda ^{k}=0,  \label{FOIG.2}
\end{equation}%
where $P\{\}$ indicates cyclic sum over the indices enclosed. Without
restriction of generality we may consider $A_{r_{1}...r_{m}}$ to be totally
symmetric; for example for two and three indices we have: 
\begin{align*}
P\left\{ A_{i;k}\right\} & =\left( A_{i;k}+A_{k;i}\right) \\
P\left\{ A_{ij;k}\right\} & =\left( A_{ij;k}+A_{kj;i}+A_{ik;j}\right).
\end{align*}%
Katzin and Levine \cite{Katzin Levine JMP 22 1981} have proved the following
results concerning First Integrals: \newpage

\begin{center}
Table 1: Collineations and corresponding First Integrals\\[0pt]

\begin{tabular}{|l|l|l|}
\hline
\textbf{Symmetry} & \textbf{Condition} & \textbf{First Integral} \\ \hline
KV & $\xi _{\left( i;j\right) }=0$ & $\xi _{i}\dot{x}^{i}$ \\ \hline
gradient KV & $\xi _{;ij}=0$ & $\xi _{;i}\dot{x}^{i}$ \\ \hline
HKV & $\xi _{i;j}=\psi g_{ij}~,~\psi _{,i}=0$ & $\xi _{i;j}\dot{x}^{i}\dot{x}%
^{j}$ \\ \hline
gradient HKV & $\xi _{;ij}=\psi g_{ij}~,~\psi _{,i}=0$ & $\xi _{;ij}\dot{x}%
^{i}\dot{x}^{j}$ \\ \hline
AC & $L_{\xi }\Gamma _{jk}^{i}=0~~,~\xi _{\left( i;jk\right) }=0$ & $\xi
_{i;j}\dot{x}^{i}\dot{x}^{j}$ \\ \hline
PC & $L_{\xi }\Gamma _{jk}^{i}=2\phi _{,(j}\delta _{k)}^{i}$ & $\left( \xi
_{i;j}-4\phi g_{ij}\right) \dot{x}^{i}\dot{x}^{j}$ \\ \hline
Special PC & $L_{\xi }\Gamma _{jk}^{i}=2\phi _{,(j}\delta _{k)}^{i}~,~\phi
_{;ij}=0$ & $\left( \xi _{i;j}-4\phi g_{ij}\right) \dot{x}^{i}\dot{x}^{j}$
\\ \hline
\end{tabular}%
\ \ \ 
\end{center}

\section{First Integrals of Noether symmetry vectors of geodesic equations}

We know that, if $X=\xi (x^{j},t)\partial _{t}+\eta ^{i}(x^{j},t)\partial
_{x^{i}}$ is the generator of a Noether symmetry with Noether function $f,$
then the quantity: 
\begin{equation}
\phi =\xi \left( \dot{x}^{i}\frac{\partial L}{\partial \dot{x}^{i}}-L\right)
-\eta ^{i}\frac{\partial L}{\partial \dot{x}^{i}}+f  \label{NS.16}
\end{equation}%
is a First Integral of $L$ which satisfies $X\phi $=0. For the Lagrangian
defined by the metric $g_{ij},$ i.e. $L=\frac{1}{2}g_{ij}\dot{x}^{i}\dot{x}%
^{j},$ we compute: 
\begin{equation}
\phi =\frac{1}{2}\xi g_{ij}\dot{x}^{i}\dot{x}^{j}-g_{ij}\eta ^{i}\dot{x}%
^{j}+f.  \label{NS.17}
\end{equation}%
In (\ref{NS.13}), (\ref{NS.14}) and (\ref{NS.15}) we have computed the
generic form of the Noether symmetry and the associated Noether function for
this Lagrangian. Substituting into (\ref{NS.17}) we find the following
expression for the generic First Integral:%
\begin{align}
\phi & =\frac{1}{2}\left[ C_{3}\psi t^{2}+2C_{2}\psi t+C_{1}\right] g_{ij}%
\dot{x}^{i}\dot{x}^{j}  \notag \\
& +\left[
C_{J}S^{J,i}+C_{I}KV^{Ii}+C_{IJ}tS^{J,i}+C_{2}H^{i}(x^{r})+C_{3}t(GHV)^{,i}%
\right] g_{ij}\dot{x}^{j}  \notag \\
& +C_{1}+C_{2}+C_{I}+C_{J}+\left[ C_{IJ}S^{J}\right] +C_{3}\left[ GHV\right]
.  \label{NS.21}
\end{align}

From the generic expression we obtain the following First Integrals\footnote{%
GHV stands for gradient HKV} :

$C_{1}\neq0.$

\begin{equation}
\phi _{C_{1}}=\frac{1}{2}g_{ij}\dot{x}^{i}\dot{x}^{j}  \label{NS.22}
\end{equation}

$C_{2}\neq 0.$%
\begin{equation}
\phi _{C_{2}}=t\psi g_{ij}\dot{x}^{i}\dot{x}^{j}-g_{ij}H^{i}\dot{x}^{j}+C_{2}
\label{NS.23}
\end{equation}

$C_{3}\neq 0.$%
\begin{equation}
\phi _{C_{3}}=\frac{1}{2}t^{2}\psi g_{ij}\dot{x}^{i}\dot{x}^{j}-t(GHV)_{,i}%
\dot{x}^{i}+\left[ GHV\right]  \label{NS.24}
\end{equation}

$C_{I}\neq 0.$%
\begin{equation}
\phi _{C_{I}}=KV_{i}^{I}\dot{x}^{i}-C_{I}  \label{NS.25}
\end{equation}

$C_{J}\neq 0.$%
\begin{equation}
\phi _{C_{J}}=g_{ij}S^{J,i}\dot{x}^{j}-C_{J}  \label{NS.26}
\end{equation}

$C_{IJ}\neq 0.$%
\begin{equation}
\phi _{IJ}=tg_{ij}S^{J,i}\dot{x}^{j}-S^{J}.  \label{NS.27}
\end{equation}

We conclude that the First Integrals of the Noether symmetry vectors of the
geodesic equations are:

a. Linear, the $\phi_{I},\phi_{J},\phi_{IJ}$

b. Quadratic, the $\phi_{c1},\phi_{c2},\phi_{3}.$

These results are compatible with the corresponding results of Katzin and
Levine \cite{Katzin Levine JMP 22 1981}.

\section{Applications}

\label{section 5}

\subsection{The Lie symmetries of geodesic equations in an Einstein space}

Suppose $X^{a}$ is a projective collineation with projection function $\phi
(x^{a}),$ such that $\mathcal{L}_{X}\Gamma _{bc}^{a}=\phi _{,b}\delta
_{c}^{a}+\phi _{,c}\delta _{b}^{a}.$ For a proper Einstein space $(R\neq 0)$
we have $R_{ab}=\displaystyle{\frac{R}{n}g_{ab}}$ from which follows:%
\begin{equation}
\mathcal{L}_{X}g_{ab}=\frac{n(1-n)}{R}\phi _{;ab}-\mathcal{L}_{X}(\ln
R)g_{ab}.  \label{NPP.2}
\end{equation}

Using the contracted Bianchi identity $\left[ R^{ij}-\frac{1}{2}Rg^{ij}%
\right] _{;j}=0$ it follows that in an Einstein space of dimension $n>2$ the
curvature scalar $R=$constant and (\ref{NPP.2}) reduces to:%
\begin{equation*}
\mathcal{L}_{X}g_{ab}=\frac{n(1-n)}{R}\phi _{;ab}.
\end{equation*}%
It follows that if $X^{a}$ generates either an affine or a special
projective collineation, then $\phi _{;ab}=0.$ Hence $X^{a}$ reduces to a
KV. This means that proper Einstein spaces do not admit HKV, ACs, special
PCs and gradient KVs (\cite{Barnes 1993}\textbf{,}\cite{YanoK})

The above results and Theorem \ref{Theorem 1} lead to the following
conclusion:

\begin{theorem}
\label{Theorem Lie Einstein space}The Lie symmetries of the geodesic
equations in a proper Einstein space of curvature scalar $R$ $\neq 0$ are
given by the vectors 
\begin{equation*}
X=\left( Kt+L\right) \partial _{t}+D^{i}\left( x\right) \partial _{i}~
\end{equation*}%
where $D^{i}(x)~$is a nongradient KV and $K,L~$are constants
\end{theorem}

\begin{theorem}
The Noether symmetries of the geodesic equations in a proper Einstein space
of curvature scalar $R$\ $\neq 0$\ are given by the vectors :\newline
\begin{equation*}
X=L\partial _{t}+D^{i}\left( x\right) \partial _{i}~~,~f=\,\text{constant}~
\end{equation*}
\end{theorem}

Theorem \ref{Theorem Lie Einstein space} extends and amends the conjecture
of \cite{Feroze Mahomed Qadir} to the more general case of Einstein spaces.

We apply the result to the case of the Euclidean 2 dimensional space. The
metric $g_{ij}$ of this space admits:

a. 2 gradient KVs, the \thinspace $Y^{1}$ and $Y^{2}$ generated by the
functions $\phi _{7},\phi _{8,}$

b. One nongradient KV, the $Y^{3},$

c. One gradient HKV, the $Y^{4}$ (with homothetic factor $1).$

Therefore the generic Noether symmetry vector is:%
\begin{equation*}
X=C_{1}\partial _{t}+C_{2}\left( 2t\partial _{t}+Y^{4}\right)
+C_{3}(t^{2}\partial
_{t}+Y^{4}t)+C_{4}tY^{1}+C_{5}tY^{2}+C_{6}Y^{1}+C_{7}Y^{2}+C_{8}Y^{3}.
\end{equation*}
where (see Table 4) 
\begin{equation*}
Y^{1}=\partial _{x},~Y^{2}=\partial _{y},~Y^{3}=y\partial _{x}-x\partial
_{y},~Y^{4}=x\partial _{x}+y\partial _{y}.
\end{equation*}

\subsection{The Noether symmetries of Schwarzschild metric}

We consider the Schwarzschild metric 
\begin{equation*}
ds^{2}=\left( 1-\frac{2m}{r}\right) dt^{2}-\frac{1}{\left( 1-\frac{2m}{r}%
\right) }dr^{2}-r^{2}\left( d\theta ^{2}+\sin ^{2}\theta d\phi ^{2}\right) .
\end{equation*}%
The geodesic Lagrangian is: 
\begin{equation*}
L=\left( 1-\frac{2m}{r}\right) \dot{t}^{2}-\frac{1}{\left( 1-\frac{2m}{r}%
\right) }\dot{r}^{2}-r^{2}\dot{\theta}^{2}-r^{2}\sin ^{2}\theta \dot{\phi}%
^{2}
\end{equation*}%
The Noether symmetries have been computed in \cite{Bokhari 2006 Int Jour
Theor Phys} as follows: 
\begin{align*}
X_{1}& =\partial _{s}~\ \ ,~X_{2}=\partial _{t}~ \\
X_{3}& =\cos \phi \partial _{\theta }-\cot \theta \sin \phi \partial _{\phi }
\\
X_{4}& =\sin \phi \partial _{\theta }+\cot \theta \cos \phi \partial _{\phi }
\\
X_{5}& =\partial _{\theta }.
\end{align*}

It easy to check that $X_{2},X_{3},X_{4}$ and $X_{5}$ are KVs of the
Schwarzschild metric and, since this metric does not admit any gradient KVs
or a HKV, these are the only Noether symmetries, a result compatible with
Theorem \ref{Theorem Noether symmetries of GEqs} above.

The first integrals of the geodesic equations of the Schwarzschild metric
are:

a. The metric integral:%
\begin{equation*}
\phi _{s}=\frac{1}{2}g_{ij}\dot{x}^{i}\dot{x}^{j}=\left( 1-\frac{2m}{r}%
\right) \dot{t}^{2}-\frac{1}{\left( 1-\frac{2m}{r}\right) }\dot{r}^{2}-r^{2}%
\dot{\theta}^{2}-r^{2}\sin ^{2}\theta \dot{\phi}^{2}
\end{equation*}

b. The linear integrals defined by the KVs:

\begin{eqnarray*}
\phi _{2} &=&\dot{t} \\
\phi _{3} &=&\left( \cos \phi \right) \dot{\theta}-\left( \cot \theta \sin
\phi ~\right) \dot{\phi} \\
\phi _{4} &=&\left( \sin \phi \right) \dot{\theta}+\left( \cot \theta \cos
\phi \right) \dot{\phi} \\
\phi _{5} &=&\dot{\theta}.
\end{eqnarray*}%
The extra Lie symmetry is $s\partial _{s}.$

\subsection{The First Integrals of the Euclidean sphere of dimension 2}

The metric of the Euclidean sphere of dimension 2 is:%
\begin{equation}
ds_{SF}^{2}=\sin ^{2}ydx^{2}+dy^{2}
\end{equation}%
The collineations admitted by the 2-d Euclidian sphere (space of constant
curvature) are three KVs and five proper PCs as shown in Table 2 \cite%
{Tsamparlis Nikolopoulos Apostolopoulos 1998}:

\begin{center}
TABLE 2: The projective algebra of the 2-d Euclidean sphere \\[0pt]
$%
\begin{tabular}{|l|l|l|}
\hline
\textbf{Type} & $\mathbf{\phi }$ & \textbf{Vector} \\ \hline
K.V & 0 & $X^{1}=\partial _{x}$ \\ \hline
K.V & 0 & $X^{2}=\cot y\cos x\partial _{x}+\sin x\partial _{y}$ \\ \hline
K.V & 0 & $X^{3}=-\cot y\sin x\partial _{x}+\cos x\partial _{y}$ \\ \hline
Pr.Col & $-\sin y\sin y\cos x$ & $X^{4}=\sin y\sin x\cos y\partial
_{x}-\left( 2\,\cos ^{2}y-1\right) \cos x\partial _{y}$ \\ \hline
Pr.Col & $-\sin y\sin y\sin x$ & $X^{5}=-\sin y\cos x\cos y\partial
_{x}-\left( 2\,\cos ^{2}y-1\right) \sin x\partial _{y}$ \\ \hline
Pr.Col & $-\sin ^{2}y\cos ^{2}x$ & $X^{6}=2\,\sin ^{2}y\cos x\sin x\partial
_{x}-2\,\cos ^{2}x\sin y\cos y\partial _{y}$ \\ \hline
Pr.Col & $-\cos x\sin x\sin ^{2}y$ & $X^{7}=\left( -2\,\cos ^{2}x+1\right)
\sin ^{2}y\partial _{x}-2\,\sin y\cos y\sin x\cos x\partial _{y}$ \\ \hline
Pr.Col & $+\cos ^{2}x\sin ^{2}y$ & $X^{8}=-2\,\sin ^{2}y\cos x\sin x\partial
_{x}-2\sin ^{2}x\sin y\cos y\partial _{y}.$ \\ \hline
\end{tabular}%
\ $
\end{center}

Hence there exist three Line First Integrals due to the KVs (dot over a
symbol indicates $\frac{d}{ds}~,~s$ being an affine parameter)\footnote{%
For example:%
\begin{equation*}
I_{1}=g_{ij}X^{i}\dot{x}^{j}=g_{11}X^{1}\dot{x}^{1}+g_{22}X^{2}\dot{x}%
^{2}=\sin ^{2}y\cdot 1\cdot \dot{x}+1\cdot 0\cdot \dot{y}=\sin ^{2}y\cdot 
\dot{x}.
\end{equation*}%
}:%
\begin{align}
I_{1}& =\dot{x}\sin ^{2}y~ \\
I_{2}& =\dot{x}\sin y\cos y\cos x~+\dot{y}\sin x~ \\
I_{3}& =-\dot{x}\sin y\cos y\sin x~+\dot{y}\cos x~
\end{align}%
and five Quadratic First Integrals corresponding to the special PCs:%
\begin{eqnarray*}
Q_{1} &=&4\dot{y}\sin x\sin ^{2}y\left( \dot{y}\cot x\cot y-\dot{x}\right) \\
Q_{2} &=&4\dot{y}\cos x\sin ^{2}y\left( \dot{x}+\dot{y}\,\tan x\tan y\right)
\\
Q_{3} &=&4\left( \dot{x}\sin y\cos x+\dot{y}\cos y\sin x\right) ^{2}-4\dot{x}%
^{2}\sin ^{2}y-4\dot{y}^{2}\cos ^{2}y \\
Q_{4} &=&4\sin x\cos x\left( \dot{x}^{2}\sin ^{2}y-\dot{y}^{2}\cos
^{2}y\right) +4\dot{x}\dot{y}\sin y\cos y\left( 1-2\cos ^{2}x\right) \\
Q_{5} &=&-4\left( \dot{x}\cos x\sin y+\dot{y}\sin x\cos y\right) ^{2}
\end{eqnarray*}

\subsection{The 1+3 decomposable metric}

We consider next the metric which is a 1+3 decomposable metric:\ 
\begin{equation}
ds_{4}=-d\tau ^{2}+U^{2}\delta _{\alpha \beta }dx^{\alpha }dx^{\beta }
\label{RW.2}
\end{equation}%
where Greek indices take the values $1,2,3.$ It is well known \cite%
{Tsamparlis Nikolopoulos Apostolopoulos 1998} that this metric admits 15
CKVs (it is conformally flat). Seven of these vectors are KVs (the six
nongradient KVs of the 3-metric $\mathbf{r}_{\mu \nu },\mathbf{I}_{\mu }$
plus the gradient KV $\partial _{\tau })$ and nine proper CKVs. The vectors
of this conformal algebra are shown in Table 3

\begin{center}
TABLE 3: The conformal algebra of the 1+3 metric (\ref{RW.2}) for $K=\pm 1$.

{\small $%
\begin{tabular}{|l|l|l|l|l|l|}
\hline
$\mathbf{K}$ & \textbf{CKVs of }$ds_{3}^{2}$ & $\mathbf{\psi }_{3}$ & 
\textbf{\#} & \textbf{CKVs of }$ds_{1+3}^{2}$ & $\mathbf{\psi }_{1+3}$ \\ 
\hline
$1$ & $H=x^{a}\partial _{a}$ & $\psi _{+}(H)=U\left( 1-\frac{1}{4}x^{\alpha
}x_{\alpha }\right) $ & 1 & $H_{1}^{+}=-\psi _{+}(H)\cos \tau \partial
_{\tau }+H\sin \tau $ & $\psi _{+}(H)\sin \tau $ \\ \hline
$1$ & $H=x^{a}\partial _{a}$ & $\psi _{+}(H)=U\left( 1-\frac{1}{4}x^{\alpha
}x_{\alpha }\right) $ & 1 & $H_{2}^{+}=\psi _{+}(H)\sin \tau \partial _{\tau
}+H\cos \tau $ & $\psi _{+}(H)\cos \tau $ \\ \hline
$1$ & $C_{\mu }=\left( \delta _{\mu }^{a}-\frac{1}{2}Ux_{\mu }x^{a}\right)
\partial _{\alpha }$ & $\psi _{+}(C_{\mu })=-Ux^{\mu }$ & 3 & $Q_{\mu
}^{+}=-\psi _{+}(C_{\mu })\cos \tau \partial _{\tau }+C_{\mu }\sin \tau $ & $%
\psi _{+}(C_{\mu })\sin \tau $ \\ \hline
$1$ & $C_{\mu }=\left( \delta _{\mu }^{\alpha }-\frac{1}{2}Ux_{\mu
}x^{a}\right) \partial _{\alpha }$ & $\psi _{+}(C_{\mu })=-Ux^{\mu }$ & 3 & $%
Q_{\mu +3}^{+}=\psi _{+}(C_{\mu })\sin \tau \partial _{\tau }+C_{\mu }\cos
\tau $ & $\psi _{+}(C_{\mu })\cos \tau $ \\ \hline
$-1$ & $H=x^{\alpha }\partial _{\alpha }$ & $\psi _{-}(H)=U\left( 1+\frac{1}{%
4}x^{\alpha }x_{\alpha }\right) $ & 1 & $H_{1}^{-}=\psi _{-}(H)\cosh \tau
\partial _{\tau }+H\sinh \tau $ & $\psi _{-}(H)\sinh \tau $ \\ \hline
$-1$ & $H=x^{\alpha }\partial _{\alpha }$ & $\psi _{-}(H)=U\left( 1+\frac{1}{%
4}x^{\alpha }x_{\alpha }\right) $ & 1 & $H_{2}^{-}=\psi _{-}(H)\sinh \tau
\partial _{\tau }+H\cosh \tau $ & $\psi _{-}(H)\cosh \tau $ \\ \hline
$-1$ & $C_{\mu }=\left( \delta _{\mu }^{\alpha }+\frac{1}{2}Ux_{\mu
}x^{a}\right) \partial _{\alpha }$ & $\psi _{-}(C_{\mu })=Ux^{\mu }$ & 3 & $%
Q_{\mu }^{-}=\psi _{-}(C_{\mu })\cosh \tau \partial _{\tau }+C_{\mu }\sinh
\tau $ & $\psi _{-}(C_{\mu })\sinh \tau $ \\ \hline
$-1$ & $C_{\mu }=\left( \delta _{\mu }^{\alpha }+\frac{1}{2}Ux_{\mu
}x^{a}\right) \partial _{\alpha }$ & $\psi _{-}(C_{\mu })=Ux^{\mu }$ & 3 & $%
Q_{\mu +3}^{-}=\psi _{-}(C_{\mu })\sinh \tau \partial _{\tau }+C_{\mu }\cosh
\tau $ & $\psi _{-}(C_{\mu })\cosh \tau $ \\ \hline
\end{tabular}%
$\\[0pt]
}
\end{center}

According to Theorem \ref{Theorem Noether symmetries of GEqs} this metric
admits the following Noether symmetries (see also \cite{Katzin Levine JMP 22
1981}): 
\begin{eqnarray*}
\partial _{s}\,~,~\mathbf{r}_{\mu \nu },\mathbf{I}_{\mu }~,\partial _{\tau
}~ &:&f=\text{constant} \\
~s\partial _{\tau }~~ &:&f=\tau \text{ }
\end{eqnarray*}%
with Noether conserved quantities:%
\begin{align}
\phi _{s}& =\frac{1}{2}g_{ij}\dot{x}^{i}\dot{x}^{j} \\
\phi _{\tau }& =\dot{\tau} \\
\phi _{\tau +1}& =s\dot{\tau}-\tau \\
\phi _{I}& =\mathbf{I}_{i}^{I}\dot{x}^{i} \\
\phi _{r}& =~\mathbf{r}_{\left( AB\right) _{i}}\dot{x}^{j}.
\end{align}

\subsection{The Noether symmetries of the FRW metrics}

In a recent paper Bokhari and Kara \cite{Bokhari 2007 Gen Rel Grav} studied
the Lie symmetries of the conformally flat Friedman Robertson Walker (FRW)
metric with the view to understand how Noether symmetries compare with
conformal Killing vectors. More specifically they considered the conformally
flat FRW\ metric\footnote{%
The second metric $ds^{2}=-t^{-\frac{4}{3}}dt^{2}+dx^{2}+dy^{2}+dz^{2}$ they
consider is the Minkowski metric whose Lie and Noether symmetries are well
known.}:

\begin{equation*}
ds^{2}=-dt^{2}+t^{\frac{4}{3}}\left( dx^{2}+dy^{2}+dz^{2}\right)
\end{equation*}%
and found that the Noether symmetries are the seven vectors:%
\begin{equation*}
\partial _{s},~S^{J}~,~r_{AB}
\end{equation*}%
where $S_{J}$ are the gradient KVs $\partial _{x}\partial _{y},\partial _{z}$
and $r_{AB}$ are the three nongradient KVs (generating SO(3)) whereas the
vector $\partial _{s}$ counts for the gauge freedom in the affine
parametrization of the geodesics. Therefore they confirm our Theorem \ref%
{Theorem Noether symmetries of GEqs} that the Noether vectors coincide with
the KVs and the HKV of the metric. Furthermore their claim that `...the
conformally transformed Friedman model admits additional conservation laws
not given by the Killing or conformal Killing vectors' \ is not correct.

In the following we compute all the Noether symmetries of the FRW\
spacetimes. To do that we have to have the homothetic algebra of these
models \cite{Maartens Maharaj CQQ 1986}. There are two cases to consider,
the conformally flat models $(K=0)$ and the nonconformally flat models ($%
K\neq 0) $.

We need the conformal algebra of the flat metric, which in Cartesian
coordinates is given in Table 4.

\begin{center}
Table 4: The conformal algebra of a flat $n-$ dimensional metric \\[0pt]
\vspace{5mm} 
\begin{tabular}{|c|c|c|c|c|c|}
\hline
\textbf{CKV} & \textbf{Components} & \textbf{\#} & $\mathbf{\psi (\xi )}$ & $%
\mathbf{F}_{ab}\mathbf{(\xi )}$ & \textbf{Comment} \\ \hline
$P_{I}$ & $\partial _{I}$ & $n$ & $0$ & $0$ & gradient KV \\ \hline
$r_{AB}$ & $2\delta _{\lbrack A}^{d}x_{B]}\partial _{d}$ & $\frac{n(n-1)}{2}$
& $0$ & $\eta _{ABab}$ & nongradient KV \\ \hline
$H$ & $x^{a}\partial _{a}$ & $1$ & $(b=1)1$ & $0$ & gradient HKV \\ \hline
$K_{I}$ & $\left[ 2x_{I}x^{d}-\delta _{I}^{d}(x_{a}x^{a})\right] \partial
_{d}$ & $n$ & $(b^{i}=\delta _{I}^{i})2x_{I}$ & $-4\eta _{I[a}x_{b]}$ & 
nongradient SCKV \\ \hline
\end{tabular}
\end{center}

\noindent Case A: $K\neq 0$

The metric is: 
\begin{equation}
ds=R^{2}\left( \tau \right) \left[ -d\tau ^{2}+\frac{1}{\left( 1+%
\displaystyle\frac{1}{4}Kx^{i}x_{i}\right) ^{2}}\left(
dx^{2}+dy^{2}+dz^{2}\right) \right] .  \label{RW.3}
\end{equation}%
For a general $R\left( \tau \right) $ this metric admits the nongradient KVs 
$~\mathbf{P}_{I},~\mathbf{r}_{\mu \nu }$ (see Table 4) and does not admit an
HKV. Therefore the Noether symmetries of the geodesic Lagrangian 
\begin{equation*}
L=-\frac{1}{2}R^{2}\left( \tau \right) \dot{\tau}^{2}+\frac{1}{2}\frac{%
R^{2}\left( \tau \right) }{\left( 1+\displaystyle\frac{1}{4}%
Kx^{i}x_{i}\right) ^{2}}\left( \dot{x}^{2}+\dot{y}^{2}+\dot{z}^{2}\right)
\end{equation*}%
of the FRW metric (\ref{RW.3}) are: 
\begin{equation*}
\partial _{s}~,~\mathbf{P}_{I}~,~\mathbf{r}_{\mu \nu }
\end{equation*}%
with Noether integrals 
\begin{equation}
\phi _{s}=\frac{1}{2}g_{ij}\dot{x}^{i}\dot{x}^{j}~,~\phi _{I}=\mathbf{P}%
_{i}^{I}\dot{x}^{i}~,~\phi _{r}=~\mathbf{r}_{\left( AB\right) _{i}}\dot{x}%
^{j}.
\end{equation}%
Concerning the Lie symmetries we note that the FRW spacetimes do not admit
ACs \textbf{\ }\cite{MaartensAC} and furthermore do not admit gradient KVs.
Therefore they do not admit special PCs. The Lie symmetries of these
spacetimes are{\large :}%
\begin{equation*}
\partial _{s},~s\partial _{s},~\mathbf{P}_{I}~,~\mathbf{r}_{\mu \nu }.
\end{equation*}

For special functions $R(\tau )$ it is possible to have more KVs and HKV. In
Table 5 we give the special forms of the scale factor $R(t)$ and the
corresponding extra KVs and HKV for $K=\pm 1$.

\newpage

\begin{center}
Table 5: The special forms of the scale factor for $K=\pm 1.$ \\[0pt]
$%
\begin{tabular}[t]{|l|l|l|l|l|l|}
\hline
$\mathbf{K}$ & \textbf{Proper CKV} & \textbf{~\#} & \textbf{Conformal Factor}
& $R(\tau )$\textbf{\ for KVs~} & $R(\tau )$\textbf{\ for HKV} \\ \hline
$\pm 1~~~~$ & $\mathbf{P}_{{\tau }}=\partial _{{\tau }}$ & $1$ & $(\ln
R\left( \tau \right) ),_{\tau }$ & $c~$ & $\exp \left( \tau \right) ~\ \ \ \
\ \ \ \ $ \\ \hline
$1$ & $\mathbf{H}_{1}^{+}$ & $1$ & -$\frac{\psi _{+}(\mathbf{H})}{R\left(
\tau \right) }\left( R\left( \tau \right) \cos \tau \right) ,_{\tau }~$ & $%
\frac{c}{\cos \tau }$ & $\nexists $ \\ \hline
$1$ & $\mathbf{H}_{2}^{+}$ & $1$ & $\frac{\psi _{+}(\mathbf{H})}{R(\tau )}%
\left( R\left( \tau \right) \sin \tau \right) ,_{\tau }$ & $\frac{c}{\sin
\tau }$ & $\nexists $ \\ \hline
$1$ & $\mathbf{Q}_{\mu }^{+}$ & $3$ & -$\frac{\psi _{+}(\mathbf{C}_{\mu })}{%
R(\tau )}\left( R\left( \tau \right) \cos \tau \right) ,_{\tau }$ & $\frac{c%
}{\cos \tau }$ & $\nexists $ \\ \hline
$1$ & $\mathbf{Q}_{\mu +3}^{+}$ & $3$ & $\frac{\psi _{+}(\mathbf{C}_{\mu })}{%
R(\tau )}\left( R\left( \tau \right) \sin \tau \right) ,_{\tau }$ & $\frac{c%
}{\sin \tau }$ & $\nexists $ \\ \hline
$-1$ & $\mathbf{H}_{1}^{-}$ & $1$ & $\frac{\psi _{-}(\mathbf{H})}{R(\tau )}%
\left( R\left( \tau \right) \cosh \tau \right) ,_{\tau }$ & $\frac{c}{\cosh
\tau }$ & $\nexists $ \\ \hline
$-1$ & $\mathbf{H}_{2}^{-}$ & $1$ & $\frac{\psi _{-}(\mathbf{H})}{R(\tau )}%
\left( R\left( \tau \right) \sinh \tau \right) ,_{\tau }$ & $\frac{c}{\sinh
\tau }$ & $\nexists $ \\ \hline
$-1$ & $\mathbf{Q}_{\mu }^{-}$ & $3$ & $\frac{\psi _{-}(\mathbf{C}_{\mu })}{%
R(\tau )}\left( R\left( \tau \right) \cosh \tau \right) ,_{\tau }$ & $\frac{c%
}{\cosh \tau }$ & $\nexists $ \\ \hline
$-1$ & $\mathbf{Q}_{\mu +3}^{-}$ & $3$ & $\frac{\psi _{-}(\mathbf{C}_{\mu })%
}{R(\tau )}\left( R\left( \tau \right) \sinh \tau \right) ,_{\tau }$ & $%
\frac{c}{\sinh \tau }$ & $\nexists $ \\ \hline
\end{tabular}%
\ $\\[0pt]
\end{center}

From Table 5 we infer the following additional Noether symmetries of the
FRW-like Lagrangian for special forms of the scale factor:

\textbf{Case A(1)}: $R\left( t\right) =c=$constant, the space is the 1+3
decomposable.

\textbf{Case A(2)} $K=1,~R\left( t\right) =\exp \left( \tau \right) .~$In
this case we have the additional gradient HKV $\mathbf{P}_{{\tau }}$
generated by the function $\frac{1}{2}\exp 2\tau $. Therefore for this scale
factor the Noether symmetry vectors, the Noether function and the conserved
Noether quantities are:~\newline
\begin{align*}
~\ ~~~~~~~~~\partial _{s}~,~\mathbf{P}_{I}~,~\mathbf{r}_{\mu \nu
}~,2s\partial _{s}+\mathbf{P}_{{\tau }}& :f=\text{constant}~~ \\
s^{2}\partial _{s}+s\mathbf{P}_{{\tau }}& :f=\frac{1}{2}\exp \left( 2\tau
\right)
\end{align*}%
with Noether Integrals 
\begin{equation*}
\phi _{s}~,~\phi _{I}~,~\phi _{r}~,~\phi _{\mathbf{\tau }}=sg_{ij}\dot{x}^{i}%
\dot{x}^{j}-g_{ij}^{i}\mathbf{P}_{{\tau }}\dot{x}^{j}~~\text{and }~\phi _{%
\mathbf{\tau +1}}=\frac{1}{2}s^{2}g_{ij}\dot{x}^{i}\dot{x}^{j}-s\mathbf{P}_{{%
\tau }}^{i}\dot{x}_{i}+\frac{1}{2}\exp 2\tau \mathbf{.}
\end{equation*}

The Lie symmetries are{\large :}%
\begin{equation*}
~\partial _{s},~s\partial _{s},~\mathbf{P}_{I}~,~\mathbf{r}_{\mu \nu }~,%
\mathbf{P}_{{\tau }},~s^{2}\partial _{s}+s\mathbf{P}_{{\tau }}\mathbf{.}
\end{equation*}

\textbf{Case A(3a)}~$K=1,~R\left( \tau \right) =\frac{c}{\cos \tau }$. In
this case we have the additional non-gradient KVs $H_{1}^{+},~Q_{\mu }^{+}.$
Therefore the Noether symmetries are:%
\begin{equation*}
\partial _{s}~,~\mathbf{P}_{I}~,~\mathbf{r}_{\mu \nu }~,H_{1}^{+},~Q_{\mu
}^{+}~~:~f=\text{constant}~
\end{equation*}%
with Noether Integrals 
\begin{equation*}
\phi _{s}~,~\phi _{I}~,~\phi _{r}~,~\phi _{H_{1}^{+}}=\left(
H_{1}^{+}\right) _{i}\dot{x}^{i}~~\text{and }~\phi _{Q_{\mu }^{+}}=\left(
Q_{\mu }^{+}\right) _{i}\dot{x}^{i}.
\end{equation*}%
The Lie symmetries are:%
\begin{equation*}
\partial _{s}~,~s\partial _{s}~,~\mathbf{P}_{I}~,~\mathbf{r}_{\mu \nu
}~,H_{1}^{+},~Q_{\mu }^{+}
\end{equation*}

\textbf{Case A(3b)}~$K=1,~R\left( \tau \right) =\displaystyle\frac{c}{\sin
\tau }.$ In this case we have the two nongradient KVs $H_{2}^{+},~Q_{\mu
+3}^{+}.$\newline
The Noether Symmetries are:%
\begin{equation*}
\partial _{s}~,~\mathbf{P}_{I}~,~\mathbf{r}_{\mu \nu }~,H_{2}^{+},~Q_{\mu
+3}^{+}~~:~f=\text{constant}~
\end{equation*}%
with Noether Integrals 
\begin{equation*}
\phi _{s}~,~\phi _{I}~,~\phi _{r}~,~\phi _{H_{2}^{+}}=\left(
H_{2}^{+}\right) _{i}\dot{x}^{i}~~\text{and }~\phi _{Q_{\mu +3}^{+}}=\left(
Q_{\mu +3}^{+}\right) _{i}\dot{x}^{i}.
\end{equation*}%
The Lie symmetries are:{\large \ }%
\begin{equation*}
\partial _{s}~,~s\partial _{s}~,~\mathbf{P}_{I}~,~\mathbf{r}_{\mu \nu
}~,H_{2}^{+},~Q_{\mu +3}^{+}.
\end{equation*}

\textbf{Case A(4a)} $K=-1,~R\left( \tau \right) =\displaystyle\frac{c}{\cosh
\tau }$ . In this case we have the two additional nongradient KVs $%
H_{1}^{-},~Q_{\mu }^{\_}.$\newline
The Noether Symmetries are:%
\begin{equation*}
\partial _{s}~,~\mathbf{P}_{I}~,~\mathbf{r}_{\mu \nu }~,H_{1}^{-},~Q_{\mu
}^{\_}~~:~f=\text{constant}~
\end{equation*}%
with Noether Integrals 
\begin{equation*}
\phi _{s}~,~\phi _{I}~,~\phi _{r}~,~\phi _{H_{1}^{-}}=\left(
H_{1}^{-}\right) _{i}\dot{x}^{i}~~\text{and }~\phi _{Q_{\mu }^{-}}=\left(
Q_{\mu }^{-}\right) _{i}\dot{x}^{i}.
\end{equation*}%
The Lie symmetries are:{\large \ }%
\begin{equation*}
\partial _{s}~,~s\partial _{s}~,~\mathbf{P}_{I}~,~\mathbf{r}_{\mu \nu
}~,H_{1}^{-},~Q_{\mu }^{\_}.
\end{equation*}

\textbf{Case A(4b)}~$K=-1,~R\left( \tau \right) =\displaystyle\frac{c}{\sin
\tau },$ we have the nongradient KV $H_{2}^{-},~Q_{\mu +3}^{\_}$.\newline
The Noether Symmetries are:%
\begin{equation*}
\partial _{s}~,~\mathbf{P}_{I}~,~\mathbf{r}_{\mu \nu }~,H_{2}^{-},~Q_{\mu
+3}^{\_}~~:~f=\text{constant}~
\end{equation*}%
with Noether Integrals 
\begin{equation*}
\phi _{s}~,~\phi _{I}~,~\phi _{r}~,~\phi _{H_{1}^{-}}=\left(
H_{2}^{-}\right) _{i}\dot{x}^{i}~~\text{and }~\phi _{Q_{\mu +3}^{-}}=\left(
Q_{\mu +3}^{-}\right) _{i}\dot{x}^{i}.
\end{equation*}%
The Lie symmetries are:{\large \ }%
\begin{equation*}
\partial _{s}~,~s\partial _{s}~,~\mathbf{P}_{I}~,~\mathbf{r}_{\mu \nu
}~,H_{2}^{-},~Q_{\mu +3}^{\_}.
\end{equation*}

\noindent \textbf{Case B}: $K=0$

In this case the metric is: 
\begin{equation*}
ds=R^{2}\left( t\right) \left( -dt^{2}+dx^{2}+dy^{2}+dz^{2}\right)
\end{equation*}%
and admits three nongradient KVs $\mathbf{P}_{I}$ and three nongradient KVs $%
\mathbf{r}_{AB}$. Therefore the Noether symmetries are:

\begin{equation*}
\partial _{s}~,\mathbf{P}_{I}~,~\mathbf{r}_{AB}~:~f=\text{constant}~
\end{equation*}%
with Noether Integrals:%
\begin{equation*}
\phi _{s}=\frac{1}{2}g_{ij}\dot{x}^{i}\dot{x}^{j}~,~\phi _{P_{I}}=\mathbf{P}%
_{i}^{I}\dot{x}^{i}~\text{and }\phi _{F}=~\left( \mathbf{r}_{AB}\right) _{i}%
\dot{x}^{j}.
\end{equation*}%
The Lie symmetries are:{\large \ }%
\begin{equation*}
\partial _{s}~,~s\partial _{s}~,~\mathbf{P}_{I}~,~\mathbf{r}_{AB}.
\end{equation*}

Again for special forms of the scale factor one obtains extra KVs and HKV as
shown in Table 6.

\begin{center}
TABLE 6: The special forms of the scale factor for $K=0$ \\[0pt]
\begin{tabular}{|l|l|l|l|l|}
\hline
\label{2RWCKV} \textbf{\#} & Proper CKV & Conformal Factor $\psi $ & $R(\tau
)$ for\ KVs~ & $R\left( \tau \right) $ for HKV \\ \hline
\textit{1} & $\mathbf{P}_{{\tau }}=\partial _{{\tau }}$ & $(\ln R(\tau ))_{,{%
\tau }}$ & $c$ & $\exp \left( \tau \right) $ \\ \hline
\textit{3} & $\mathbf{M}_{{\tau \alpha }}=x_{{\alpha }}\partial _{{\tau }}+{%
\tau }\partial _{{\alpha }}$ & $x_{{\alpha }}(\ln R(\tau ))_{,{\tau }}$ & $c$
& $\nexists $ \\ \hline
\textit{1} & $\mathbf{H}=\mathbf{P}_{{\tau }}+x^{a}\partial _{a}$ & $\mathbf{%
\tau }(\ln R(\tau ))+1$ & $c/\tau $ & $\nexists $ \\ \hline
\textit{1} & $\mathbf{K}_{{\tau }}=2{\tau }\mathbf{H}+\left( x^{c}x_{c}-\tau
^{2}\right) \partial _{{\tau }}$ & $-(\ln R(\tau ))_{,\tau }\left( -\tau
^{2}+x^{2}+y^{2}+z^{2}\right) +2\epsilon \tau $ & $\nexists $ & $\nexists $
\\ \hline
\textit{3} & $\mathbf{K}_{\mu }=2x_{\mu }\mathbf{H}-\left( x^{c}x_{c}-\tau
^{2}\right) \partial _{\mu }~$ & $2x_{\mu }\left[ \tau (\ln R(\tau ))_{,\tau
}+1\right] $ & $c/\tau $ & $\nexists $ \\ \hline
\end{tabular}
\end{center}

From Table 6 we have the following special cases.

\textbf{Case B(1)}: $R\left( t\right) =c=$constant. Then the space is the
Minkowski space.

\textbf{Case B(2):} $R\left( t\right) =\exp \left( \tau \right) .$ Then $%
\mathbf{P}_{{\tau }}~$becomes a gradient HKV $\left( \psi =1,~\text{gradient
function}~\frac{1}{2}\exp \left( 2\tau \right) \right) .$ Hence the Noether
symmetries are%
\begin{align*}
~~\ \ \ \ \ \ \ \ \ \partial _{s}~,\mathbf{P}_{I}~,~\mathbf{r}%
_{AB}~,~2s\partial _{s}+\mathbf{P}_{{\tau }}~~& :~f=\text{constant}~ \\
s^{2}\partial _{s}+s\mathbf{P}_{{\tau }}~~& :~f=\frac{1}{2}\exp \left( 2\tau
\right)
\end{align*}%
with Noether Integrals 
\begin{equation*}
\phi _{s}~,~\phi _{P_{I}}~,~\phi _{F}~,~\phi _{\mathbf{P}_{{\tau }}}=sg_{ij}%
\dot{x}^{i}\dot{x}^{j}-g_{ij}\left( \mathbf{P}_{{\tau }}\right) ^{i}\dot{x}%
^{j}~~\text{and }~\phi _{\mathbf{Y+1}}=\frac{1}{2}s^{2}g_{ij}\dot{x}^{i}\dot{%
x}^{j}-s\left( \mathbf{P}_{{\tau }}\right) _{i}\dot{x}^{i}+\mathbf{P}_{{\tau 
}}.
\end{equation*}%
The Lie symmetries are:{\large \ }%
\begin{equation*}
\partial _{s}~,~s\partial _{s}~,~~\mathbf{P}_{I}~,~\mathbf{r}_{AB}~,~\mathbf{%
P}_{{\tau }}~,s^{2}\partial _{s}+s\mathbf{P}_{{\tau }}.~
\end{equation*}

\textbf{Case B(3):} $R\left( t\right) =\tau ^{-1}~.$ Then $\ $we have four
additional nongradient KVs, the $\mathbf{H},$ and $\mathbf{K}_{\mu },$ and
the Noether symmetries are: 
\begin{equation*}
\partial _{s}~,~\mathbf{P}_{I},~\mathbf{r}_{AB}~,~\mathbf{H~,~K}_{\mu }:~f=%
\text{constant}~~
\end{equation*}%
with Noether Integrals 
\begin{equation*}
\phi _{s}~,~\phi _{P_{I}}~,~\phi _{F}~,~\phi _{~\mathbf{H}}=\left( \mathbf{H}%
\right) _{i}\dot{x}^{j}~~\text{and }~\phi _{\mathbf{K}_{\mu }}=\left( 
\mathbf{K}_{\mu }\right) _{i}\dot{x}^{i}.
\end{equation*}%
The Lie symmetries are:{\large \ }%
\begin{equation*}
\partial _{s}~,~s\partial _{s}~,~\mathbf{P}_{I},~\mathbf{r}_{AB}~,~\mathbf{%
H~,~K}_{\mu }.
\end{equation*}

\section*{Acknowledgements}

One of the authors (MT) would like to express his sincere gratitude to
Professor Peter Leach who, a few years back during a visit at the Faculty of
Physics of the University of Athens, introduced him to the topic of Lie
symmetries of differential equations and motivated the present work.


\begin{thebibliography}{99}
\bibitem{Katzin Levine 1972 Poland} Katzin G.H., Levine J.: Colloquium
Mathematicum (Wroc%
l{\hskip.15em}\llap{\raise.25ex\hbox{$\cdot$}}%
aw, Poland) 21 (1972),

\bibitem{Katzin  Levine JMP 15 1974} Katzin G.H., Levine J.: J. Math. Phys 
\textbf{15},\ 1460 (1974)

\bibitem{Katzin  Levine JMP 17  1976} Katzin G.H., Levine J.: J. Math. Phys\ 
\textbf{17}, 1345 (1976)

\bibitem{Katzin Levine JMP 22 1981} Katzin G.H., Levine J.:\emph{\ }J. Math.
Phys\ \textbf{22},1878 (1981)

\bibitem{Aminova 1978} Aminova, A.V.: Gravit.i. Teoriya Otmotisel' \textbf{14%
}, 4 (1978)

\bibitem{Aminova 1994} Aminova, A.V.: Izv. - Vyssh. - Uchebn. - Zaved. -
Mat. \textbf{2}, 3 (1994)

\bibitem{Aminova 1995} Aminova, A.V.: Sbornik Mathematics \textbf{186} (12),
1711 (1995)

\bibitem{Aminova 2000} Aminova, A.V.: Tensor, N.S., \textbf{65} (2000)

\bibitem{Prince Crampin (1984) 1} Prince G. E., Crampin M.: \emph{\ }Gen Rel
Grav. \textbf{16}, 921 (1984)

\bibitem{Feroze Mahomed Qadir} Feroze, T., Mahomed, F.M., Qadir, A.: \emph{\ 
}Nonlinear Dynamics 45, 65 (2006)

\bibitem{Bokhari 2006 Int Jour Theor Phys} Bokhari A.H., Kara A.H., Kashif
A.R., Zaman F. D.: Inter. Jour. Theor. Phys.\textbf{\ 45}, 1063 (2006)

\bibitem{Bokhari 2007 Gen Rel Grav} Bokhari A.H., Kara A.H.: Gen Rel Grav. 
\textbf{39}, 2053 - 2059 (2007)

\bibitem{Yano 1956} Yano K.: The theory of Lie derivatives and its
Applications. Amsterdam: North Holland (1956)

\bibitem{Scouten 1954} Scouten K.J.A.: Ricci Calculus, Springer (1954)

\bibitem{Stephani book ODES} Stephani H.: Differential Equations: Their
Solutions using Symmetry, Cambridge University Press, NY (1989)

\bibitem{Tsamparlis Nikolopoulos Apostolopoulos 1998} Tsamparlis, M.,
Nikolopoulos, D., Apostolopoulos, P. S.: Class. Quantum Grav. \textbf{15},
2909 (1998)

\bibitem{Barnes 1993} Barnes A.:\ Class. Quantum Grav. \textbf{10}, 1139
(1993)\newline

\bibitem{YanoK} Knebelman M.S., Yano K.,: Proc. Amer. Math. Soc., \textbf{12}%
, 300 (1961)

\bibitem{Maartens Maharaj CQQ 1986} Maartens R., Maharaj S.D.: Class Quant
Grav. \textbf{3}, 1005 (1986)

\bibitem{MaartensAC} Maartens R.: J. Math. Phys 28, 2051 (1987)
\end{thebibliography}
\end{document}